\documentclass[a4paper,UKenglish,cleveref, autoref, thm-restate,authorcolumns]{lipics-v2019}

\newif\ifDoubleBlind
\DoubleBlindfalse

\newif\ifTR
\TRtrue
\usepackage{pgfplots}
\usepackage{tikz}
\usetikzlibrary{matrix,fit}
\usepackage{amsfonts}

\newcommand{\set}[1]{\left\{ #1\right\}}
\newcommand{\gilt}{:}
\newcommand{\sodass}{\,:\,}
\newcommand{\setGilt}[2]{\left\{ #1\sodass #2\right\}}

\newcommand{\defeq}{\mathrel{:=}}

\newcommand{\realrange}[2]{\left[#1, #2\right]}

\newcommand{\unitrange}[2]{\realrange{0}{1}}

\newcommand{\Ohh}[1]{\mathcal{O}\!\left( #1\right)}

\newcommand{\llabel}[1]{\label{\labelprefix:#1}}
\newcommand{\labelprefix}{} % later redefined using renewcommand

\newcommand{\discussionsize}{\small}

\newcommand{\frage}[1]{}

\newenvironment{code}{\noindent%\sf%
\begin{tabbing}%
\hspace{2em}\=\hspace{2em}\=\hspace{2em}\=\hspace{2em}\=\hspace{2em}\=%
\hspace{2em}\=\hspace{2em}\=\hspace{2em}\=\hspace{2em}\=\hspace{2em}\=%
\kill}{\end{tabbing}}

\newcommand{\labelcommand}{}
\newcommand{\captiontext}{}
\newsavebox{\codeparam}
\newcounter{lineNumber}
\newenvironment{disscodepos}[3]{%
\renewcommand{\labelcommand}{#2}%
\renewcommand{\captiontext}{#3}%
\sbox{\codeparam}{\parbox{\textwidth}{#3}}%
\begin{figure}[#1]\begin{center}\begin{code}\setcounter{lineNumber}{1}}{%
\end{code}\end{center}\caption{\llabel{\labelcommand}\captiontext}\end{figure}}

{\end{disscodepos}}

\newcommand{\Is}       {:=}

\newcommand{\Return}   {{\bf return\ }}

\newdimen\endofsize\endofsize=0.5em
\def\endofbeweis{~\quad\hglue\hsize minus\hsize
                 \hbox{\vrule height \endofsize width
\endofsize}\par}
\usepackage{amsmath}
\usepackage{amssymb}
\usepackage[ruled,vlined,linesnumbered,norelsize]{algorithm2e}
\DontPrintSemicolon
\DontPrintSemicolon

\let\oldnl\nl \newcommand\nonl{\renewcommand{\nl}{\let\nl\oldnl}}
\usepackage{booktabs}
\usepackage{numprint}
\npdecimalsign{.} % we want . not , in numbers

\usepackage{float}

\usepackage[numbers,square,sort&compress]{natbib}

\newcommand{\eg}{e.g.\ }

\newcounter{red}
\newtheorem{reduction}[red]{Reduction}
\newtheorem{inexactreduction}[red]{Reduction}

\DeclareMathOperator*{\argmin}{arg\,min}

\usepackage[draft]{changes}
\definechangesauthor[name={Darren Strash}, color=blue]{DS} 
\definechangesauthor[name={Wolfgang Ost}, color=green]{WO}

\ifTR 
\newcommand{\new}[1]{{#1}}
\else 
\newcommand{\new}[1]{{#1}}
\fi

\EventLongTitle{}
\EventShortTitle{}
\EventAcronym{}
\EventYear{}
\EventDate{}
\EventLocation{}
\EventLogo{}
\nolinenumbers
\hideLIPIcs

\title{Engineering Data Reduction for Nested Dissection} 

\titlerunning{Engineering Data Reduction for Nested Dissection}

\ifDoubleBlind
\author{B. Anonymous}{A University}{}{}{}
\authorrunning{B. Anonymous}
\Copyright{The Authors}
\else
\author{Wolfgang Ost}{ Faculty of Computer Science, University of Vienna, Austria}{wolfgang.ost@univie.ac.at}{https://orcid.org/0000-0003-4311-9928}{}

\author{Christian Schulz}{ Faculty of Computer Science, University of Vienna, Austria}{christian.schulz@univie.ac.at}{https://orcid.org/0000-0002-2823-3506}{Partially supported by DFG grant SCHU 2567/1-2.}

\author{Darren Strash}{ Department of Computer Science, Hamilton College, USA}{dstrash@hamilton.edu}{https://orcid.org/0000-0001-7095-8749}{}

\authorrunning{W. Ost and C. Schulz and D. Strash}

\fi{}


\category{} 

\relatedversion{}
\supplement{}

\acknowledgements{}%optional

\EventEditors{John Q. Open and Joan R. Access}
\EventNoEds{2}
\EventLongTitle{42nd Conference on Very Important Topics (CVIT 2016)}
\EventShortTitle{CVIT 2016}
\EventAcronym{CVIT}
\EventYear{2016}
\EventDate{December 24--27, 2016}
\EventLocation{Little Whinging, United Kingdom}
\EventLogo{}
\SeriesVolume{42}
\ArticleNo{23}
\begin{document}

\maketitle

\begin{abstract}
Many applications rely on time-intensive matrix operations, such as factorization, which can be sped up significantly for large sparse matrices by interpreting the matrix as a sparse graph and computing a node ordering that minimizes the so-called \emph{fill-in}. 
In this paper, we engineer new data reduction rules for the minimum fill-in problem, which significantly reduce the size of the graph while producing an equivalent (or near-equivalent) instance. 
By applying both new and existing data reduction rules exhaustively before nested dissection, we obtain improved quality \emph{and} \new{at the same time large improvements} in running time on a variety of instances.
Our overall algorithm outperforms the state-of-the-art significantly:
it not only yields better elimination orders, but it does so significantly faster than previously possible.
 \new{For example, on road networks, where nested dissection algorithms are typically used as a preprocessing step for shortest path computations, our algorithms are on average six times faster than Metis while computing orderings with less fill-in.}
\end{abstract}
\setcounter{page}{0}
\section{Introduction}
Solving sparse linear systems of equations
is a fundamental task in scientific computing with a variety of applications, such as computational fluid dynamics, electrical flows, structural engineering, economic modeling and circuit simulation~\cite{davis11,DBLP:conf/icalp/Spielman12}. \new{Another important application is solving Laplacian systems which is, among many other use cases, needed to gain insights on the spectral properties of a given network by examining the eigenvalues and eigenvectors of the graph Laplacian~\cite{DBLP:conf/icalp/Spielman12}.} 
\emph{Sparse linear systems}, of the form $Ax = b$, can in principle be solved by direct methods~\cite{golub13,davis16}. Such methods decompose the matrix $A$ into factors that simplify solving  the system. The drawback is that such factors can become dense, having many more non-zeros than the original matrix~\cite{rose72,golub13,davis16}. Solving the system then becomes prohibitively expensive in terms of storage and computation time. The number of new non-zeros introduced by factorization is called the \emph{fill-in}.  By reordering the system, fill-in can be significantly reduced, leading to sparse factors~\cite{rose72,davis16,golub13}. Thus, a problem of central importance is to reduce the fill-in as much as possible to reduce computation time and storage overhead.
For symmetric positive definite matrices, which can be factored by Cholesky factorization~\cite{golub13}, we can reorder rows and columns by a symmetric permutation $PAP^\top$~\cite{rose72,golub13}. The minimum fill-in problem is to find a permutation matrix $P$, such that the number of non-zeros introduced during factorization is minimized.

However, Yannakakis~\cite{yannakakis81} has shown that the problem is NP-complete.
Hence, heuristic algorithms such as the minimum degree algorithm~\cite{tinney67,rose72}, nested dissection~\cite{george73} or combinations of both that work on a graph representation of the input matrix are typically used in practice. 
More precisely, a symmetric matrix can be represented by an undirected graph. In this graph nodes represent rows and columns of the matrix. There is an edge $\{u, v\}$ in the graph if the matrix element $a_{u,v}$ is not zero. An elimination step in the matrix is reflected in the graph by removing the node corresponding to the eliminated column and connecting its neighborhood to form a clique. The added edges provide an upper bound to the number of non-zeros introduced in an elimination step.

On the other hand, many NP-hard graph problems have been shown to be fixed-parameter tractable (FPT): large inputs can be solved efficiently and provably optimally, as long as some parameter of the input is small.  Over the last two decades, significant advances have been made in the design and analysis of FPT algorithms for a wide variety of graph problems. 
This has resulted in a rich algorithmic toolbox that is by now well-established and described in several textbooks and surveys, \eg \cite{DBLP:books/sp/CyganFKLMPPS15,DBLP:journals/eatcs/Kratsch14}. 
Few of the new techniques are implemented and tested on real datasets, and their practical potential is far from understood. 
\ifTR
However, recently the engineering part in this area has gained some momentum. 
For example, three years ago the Parameterized Algorithms and Computational Experiments Challenge (PACE)~\cite{dell_et_al:LIPIcs:2018:8558} was introduced. Here teams compete to solve real-world inputs using ideas from parameterized algorithm design. More than 75 researchers from 16 countries participated in the 2018 iteration of the challenge. PACE continues to bring together theory and practice in the parameterized algorithms community.
In addition there are several experimental studies in the area that take up ideas from FPT or kernelization theory, \eg for independent sets (or equivalently vertex cover)~\cite{akiba-tcs-2016,DBLP:conf/sigmod/ChangLZ17,dahlum2016accelerating,DBLP:conf/alenex/Lamm0SWZ19,DBLP:journals/corr/abs-1908-06795,DBLP:conf/alenex/Hespe0S18}, for cut tree construction\cite{DBLP:conf/icdm/AkibaISMY16}, for matching \cite{DBLP:conf/esa/KorenweinNNZ18}, for treewidth computations \cite{bannach_et_al:LIPIcs:2018:9469,DBLP:conf/esa/Tamaki17,koster2001treewidth}, for the feedback vertex set problem \cite{DBLP:conf/wea/KiljanP18,DBLP:conf/esa/FleischerWY09}, for the dominating set problem~\cite{10.1007/978-3-319-55911-7_5}, for the minimum cut~\cite{DBLP:conf/alenex/HenzingerN0S18,DBLP:conf/ipps/HenzingerN019}, for multiterminal cut problem~\cite{henzinger2019sharedmemory}, for the maximum cut problem~\cite{DBLP:journals/corr/abs-1905-10902} and for the cluster editing problem~\cite{Boecker2011}.
Surprisingly, the minimum fill-in problem also admits a wide range of simple data reduction techniques that have not yet been successfully used in practice. 
\else 
However, recently the engineering part in area has gained some momentum~\cite{DBLP:conf/alenex/Lamm0SWZ19,DBLP:journals/corr/abs-1908-06795,DBLP:conf/alenex/Hespe0S18,DBLP:conf/icdm/AkibaISMY16,DBLP:conf/esa/Tamaki17,henzinger2019sharedmemory,DBLP:conf/alenex/HenzingerN0S18,DBLP:conf/ipps/HenzingerN019,dahlum2016accelerating}. 
Surprisingly, the minimum fill-in problem also admits a wide range of simple data reduction techniques that have not yet been successfully used in practice. 
\fi

\textbf{Our Results.}
We engineer a new node ordering algorithm that employs novel and existing data reduction rules before using a nested dissection algorithm. After the nested dissection algorithm terminates, reductions are undone to compute the final node ordering.
By applying data reduction rules exhaustively we obtain improved quality \emph{and} \new{at the same time large improvements} in running time on a variety of instances.
Overall, we arrive at a system that outperforms the state-of-the-art significantly.
 \new{For example, on road networks, where nested dissection algorithms are typically used as a preprocessing step for shortest path computations~\cite{DBLP:journals/jea/DibbeltSW16,DBLP:journals/algorithms/GottesburenHUW19}, our algorithms are on average six times faster than Metis while computing orderings with less fill-in.} \ifTR\else (If accepted, we will make the implementation freely available.)\fi{}

\section{Preliminaries}
\label{s:preliminaries}
In the following we consider an undirected graph $G=(V,E)$, where $V$ are the vertices and $E$ are the edges.
We use $|V| = n$ and $|E| = m$. 
$\Gamma_G(v) \Is \setGilt{u}{\set{v,u}\in E}$ denotes the \emph{neighborhood} of a node $v$.
The set $\Gamma_G[v] \Is \Gamma_G(v) \cup\{v\}$ is the \emph{closed neighborhood} of $v$ in $G$.
For a set of nodes $A \subseteq V$ we define its neighborhood $\Gamma_G(A) \defeq \left(\bigcup_{x \in A} \Gamma_G(x)\right) \setminus A$.
When clear from the context we omit $G$ and write $\Gamma(x)$, $\Gamma[x]$ and $\Gamma(A)$, respectively.

For a set of nodes $V' \subseteq V$ we define the set of edges with both endpoints in $V'$ as $E(V') \defeq E \cap (V' \times V')$.
A graph $S=(V', E')$ is said to be a \emph{subgraph} of $G=(V, E)$ if $V' \subseteq V$ and $E' \subseteq E(V')$.
We call $S$ an \emph{induced} subgraph when $E' = E(V')$.  
For a set of nodes $U\subseteq V$, $G[U]$ denotes the subgraph induced by $U$.

A graph $G$ is \emph{triangulated} or \emph{chordal}, if for every cycle of four or more nodes, there is an edge connecting two non-consecutive nodes in the cycle.
A \emph{triangulation} of a graph $G = (V, E)$ is a set of edges $T$, such that $(V, E \cup T)$ is a triangulated graph.
A triangulation is \emph{minimal} if no proper subset is also a triangulation.
If there is no triangulation $T'$ with $|T'| < |T|$, then $T$ is a \emph{minimum triangulation}.
A \emph{clique} is a set of vertices $K\subseteq V$ such that $\forall u,v\in K$ where $u\neq v$\,\, $\{u,v\}\in E$. A vertex $v\in V$ is \emph{simplicial} if $\Gamma(v)$ is a clique. A graph $G$ is said to have a \emph{perfect elimination ordering} if there is an ordering of vertices $v_1v_2\cdots v_n$ such that each vertex $v_i$ is simplicial in the subgraph~$G[\{v_{i+1},\ldots,v_{n}\}]$ induced by vertices later in the ordering.

In this work, we consider several related partitioning problems.
The \emph{graph partitioning problem} asks for \emph{blocks} of nodes $V_1$,\ldots,$V_k$ 
that partition $V$; that is, $V_1\cup\cdots\cup V_k=V$ and $V_i\cap V_j=\emptyset$
for $i\neq j$. A \emph{balancing constraint} demands that 
$\forall i\in \{1..k\}\gilt |V_i|\leq L_{\max}\Is (1+\epsilon)\lceil |V|/k \rceil$ for
some parameter $\epsilon$. 
In this case, the objective is often to minimize the total \emph{cut} $\sum_{i<j}|E_{ij}|$ where 
$E_{ij}\Is\setGilt{\set{u,v}\in E}{u\in V_i,v\in V_j}$. 
The set of cut edges is also called an \emph{edge separator}.
A node $v \in V_i$ that has a neighbor $w \in V_j, i\neq j$, is a \emph{boundary node}.
The \emph{node separator problem} asks to find blocks, $V_1, V_2$ and a separator $S$ that partition $V$ such that there are no edges between the blocks. 
Again, a balancing constraint demands $|V_i| \leq (1+\epsilon)\lceil|V|/k \rceil $. However, there is no balancing constraint on the separator $S$. 
The objective is to minimize the size of the separator $|S|$. 
We call $V_1$ and $V_2$ the \emph{components} and the induced subgraphs $G[S \cup V_i]$ the \emph{leaves} of $S$. A separator that is also a clique is a \emph{separation clique}.

A  \emph{multilevel approach} consists of three main phases: coarsening, initial solution, and uncoarsening.
In the coarsening phase, 
contraction should quickly reduce the size of the input.
Contraction is stopped when the graph is small enough so a problem can be solved by some other potentially more expensive algorithm, producing the initial solution. 
In the uncoarsening phase, contractions are iteratively undone and local search is used on all levels to improve a solution.  
The intuition behind the approach is that a good solution at one level of the hierarchy will also be a good solution on the next finer level so that local search will quickly find a good solution.

\textbf{The Node Ordering Problem.}
Given a matrix $A \in \mathbb{R}^{n\times n}$ and a column vector $b \in \mathbb{R}^n$ we want to solve the system of linear equations given by $Ax=b$. This is usually accomplished by first factoring the matrix $A$. For symmetric matrices the Cholesky decomposition can be used which factorizes $A$ into a lower triangular matrix $L$ and its transpose $L^\top$ such that $A=LL^\top$. An extension of the simple Cholesky decomposition is to reorder the rows and columns of $A$ prior to the factorization. This is done by applying a permutation matrix $P$ to rows and columns of the matrix $A$ which leads to $P A P^\top = LL^\top$. 
For large sparse matrices it is crucial to choose a good permutation matrix $P$ in order to reduce the fill-in during the factorization which reduces both the amount of memory needed to store the factors as well as the number of operations needed to factorize the matrix. A permutation matrix can also be expressed as a permutation vector which maps each row respectively column to a rank in $\{1,\dots,n\}$.
The matrix $A$ can be viewed as a graph $G=(V,E)$ such that $V := \{1,\dots,n\}$ and there exists an edge for every non-zero entry in $A$ which does not lie on the diagonal: $E := \{\{i,j\} : i \neq j \wedge A[i,j] \neq 0\}$. Elimination of a column and row in $A$ is reflected in $G$ by eliminating the corresponding node and connecting its neighborhood to form a clique. Finding a permutation matrix for $A$ then corresponds to finding an elimination order of nodes in $G$, which is called a \emph{node ordering}.

The \emph{deficiency} $D_G(x)$ of a node $x$ in a graph $G$ is the set of distinct pairs of nodes in $\Gamma_G(x)$, that are not themselves neighbors:
       $ D_G(x) := \{\{a, b\} \mid a, b \in \Gamma_G(x),\ a \neq b,\ a \notin \Gamma_G(b) \}$.
When clear from the context we omit $G$ and write $D(x)$.
Eliminating a node $x$ from a graph $G = (V, E)$ results in the \emph{elimination graph} $G_x := (V \setminus \{x\}, E(V \setminus \{x\}) \cup D_G(x))$, which is obtained by removing $x$ and its incident edges from $G$, and connecting the neighborhood of $x$ to a clique.
The elimination graph obtained by eliminating a sequence of nodes $X = x_1 x_2 \cdots x_m$ is denoted by 
        $G_X := (\dotsc((G_{x_1})_{x_2})\dotsc)_{x_m}$.

        A node ordering of a graph $G = (V, E)$ with $n = |V|$ is a bijection $\sigma: \{1,2,\dotsc,n\} \to V$, that defines a sequence of elimination graphs $G^{(1)} G^{(2)} \dotsc G^{(n)}$, where
        $G^{(i)} :=
                (G^{(i-1)})_{\sigma(i)}         \text{ if } i = 1,\dotsc,n \text{ and } 
                G                               \text{ if } i = 0$.
In $G^{(n)}$, all nodes have been eliminated. 
The \emph{fill-in} of an ordering $\sigma$ is the number of edges added during the elimination process, denoted by
        $\phi(G, \sigma) := \sum_{i = 1}^{n} |D_{G^{(i-1)}}(\sigma(i))|$.
We let $\Sigma(G) = \argmin_\sigma \{\phi(G, \sigma)\}$ be some minimum fill-in ordering of a graph $G$, with the corresponding minimum fill-in $\Phi(G) = \phi(G, \Sigma(G))$. Note that
\begin{equation}
        \Phi(G) \geq \Phi(G^{(1)}) \geq \dotsc \geq \Phi(G^{(n-1)}).
        \label{eq:phi-non-decreasing}
\end{equation}

An ordering $\sigma$ of a graph $G = (V,E)$ generates a triangulation $T(\sigma)$ of G, such that the graph $(V, E \cup T(\sigma))$ is chordal. $T(\sigma)$ is the set of edges added during the elimination process and $|T(\sigma)| = \phi(G, \sigma)$.
A minimum fill-in ordering $\Sigma(G)$ generates a \emph{minimum triangulation} $T(\Sigma(G))$, where $\Phi(G) = |T(\Sigma(G))|$~\cite{ohtsuki76}. If $G$ is triangulated, then its minimum triangulation is the empty set and it has a perfect elimination order, i.e., $\Phi(G) = 0$.

We use the following notation for node orderings:
        $\sigma = x_1 x_2 \cdots x_n$
corresponds to
        $\sigma(1) = x_1, \sigma(2) = x_2, \dots, \sigma(n) = x_n$.
We write $x \Sigma(G_x)$ if $x$ is to be eliminated before the nodes in $G_x$. To denote nodes ordering where a set of nodes $P = \{p_1,p_2,\dotsc,p_n\}$ are eliminated in any order, we use $P$ in the notation instead of $p_1 p_2 \cdots p_n$. For example, $P \Sigma(G_P)$ is an ordering in which the nodes in $P$ are eliminated in any order before the nodes in $G_P$.

%%%%%%%%%%%%%%%%%%%%%%%%%%%%%%%%%%%%%%%%%%%%%
\section{Related Work}
\label{s:related}
There has been a \emph{huge} amount of research on graph partitioning, node separators and minimum fill-in ordering; we refer the reader to the overviews~\cite{GPOverviewBook,SPPGPOverviewPaper,DBLP:reference/bdt/0003S19} for preliminary material in this area. 
Here, we focus on issues closely related to our main contributions and previous work on the node ordering problem. 

Yannakakis proved that the problem of finding a minimum fill-in ordering is NP-complete~\cite{yannakakis81}. Exact algorithms have been introduced in the context of non-serial dynamic programming~\cite{bertele69.1,bertele69.2}, but they are not practical for large matrices due to their exponential running time~\cite{rose72}. For graphs with a perfect elimination order, the problem can be solved in $\Ohh{|V| + |E|}$ time~\cite{rose76}.
Tinney and Walker~\cite{tinney67} introduced a heuristic algorithm where the next column to eliminate is selected based on the number of non-zeros. This algorithm is known as the \emph{minimum degree algorithm}, since a node of minimum degree is eliminated at each step~\cite{rose72}. There have been several improvements to this algorithm, both in its design and implementation~\cite{george78, george80, george89}.
The minimum degree algorithm spends a significant part of its time in updating node degrees. Most of the improvements to the minimum degree algorithm are thus focused on reducing the number of nodes to update~\cite{george89}. Amestoy et al.~\cite{amestoy96} introduced an approximate minimum degree algorithm in which the degree update is not performed exactly.
The \emph{minimum deficiency algorithm} is a greedy algorithm similar to the minimum degree algorithm~\cite{rose72,tinney67}: at every step the node with the smallest deficiency is eliminated. If the graph to be ordered has a perfect elimination ordering, the minimum deficiency algorithm finds it. However, finding the deficiency of a node is expensive, so the algorithm is slower than the minimum degree algorithm \cite{rose72}.

In 1973, George \cite{george73} introduced an algorithm to produce orderings for regular finite element meshes, called nested dissection. This algorithm computes a node separator, and then recursively orders the partitions before the separator. George and Liu generalized the algorithm to work on arbitrary graphs \cite{george78.2}. In practice, nested dissection is combined with algorithms such as the minimum degree algorithm: once the subgraphs are small enough, they are ordered by the minimum degree algorithm~\cite{ashcraft94,ashcraft98,karypis98}. A similar approach based on multisectors instead of bisectors was presented by Ashcraft and Liu~\cite{ashcraft98}.
LaSalle and Karypis~\cite{lasalle2015efficient} gave a shared-memory parallel algorithm to compute node separators used to compute fill-reducing orderings.
Within a multilevel approach they evaluate different local search algorithms indicating that a combination of greedy local search with a segmented FM algorithm can outperform serial FM algorithms. 
\new{On road networks nested dissection is used as preprocessing step for shortest path computations~\cite{DBLP:journals/algorithms/GottesburenHUW19}. The authors use degree-2 preprocessing to speed up their nested dissection algorithm.}

\section{Advanced Node Ordering}
\label{sec:reductions}
\label{s:mainsection}

We now outline our reduced nested dissection algorithm and describe our reductions in detail.
For completeness, we outline the standard nested dissection algorithm in Algorithm \ref{alg:reduced-nested-dissection} as implemented for example in Metis~\cite{karypis1998fast}.
We extend the nested dissection by transforming the input graph $G$ into a (smaller) equivalent graph $G'$ using our reduction rules.
We apply reductions in a fixed order and each reduction is applied exhaustively, i.e., the graph is reduced as much as possible by each reduction.
Then, we apply nested dissection on the reduced graph $G'$ to obtain an ordering $\sigma$.
After the nested dissection algorithm returns the ordering $\sigma$, the ordering of the reduced graph is then transformed to an ordering of the input graph~$\sigma'$. We now explain the data reduction rules that we use.

\ifTR
\begin{algorithm}[t]
        \SetKwFunction{Separator}{Separator}
        \SetKwFunction{MinDegree}{MinDegree}
        \SetKwFunction{NestedDissection}{UnreducedNestedDissection}
        \SetKwFunction{ReduceGraph}{ReduceGraph}
        \SetKwFunction{MapOrdering}{MapOrdering}
        \SetKwInOut{Input}{input}\SetKwInOut{Output}{output}
        \SetKw{in}{in}

        \Input{Undirected graph $G = (V, E)$}
        \Output{Ordering $\sigma$}

        \BlankLine

        \eIf {$|G| \geq \text{recursion limit}$} {
                $V_1, V_2, S \leftarrow$ \Separator{$G$}\;
                \ForEach{$G'$ \in $(G[V_1], G[V_2], G[S])$}{
                        $\sigma' \leftarrow$ \NestedDissection{$G'$}\;
                        $\sigma \leftarrow \sigma\sigma'$\;
                }
        } { $\sigma \leftarrow$ \MinDegree{G}\; }
        \Return $\sigma$

        \caption{UnreducedNestedDissection($G$)}
        \label{alg:reduced-nested-dissection}
\end{algorithm}
\fi{}

\subsection{Data Reduction Rules}
A \emph{data reduction rule} transforms an input graph $G$ into a smaller, reduced graph $G'$. This new smaller problem instance is generally equivalent to the original, and can be solved in less time. The solution on $G'$ can then be transformed into a node ordering of the nodes of $G$. If the running time overhead of these transformations is sufficiently small, solving the problem on $G$ in this way will be faster than a direct approach.

We use four exact and two inexact reduction rules. The \emph{simplicial node reduction} eliminates nodes whose neighborhood is already a clique. These nodes can be ordered first in a minimum fill-in ordering, since they do not contribute to the fill-in. The \emph{indistinguishable node reduction} and \emph{twin reduction} contract sets of nodes with equal closed and open neighborhood, respectively. When any node in such a set is eliminated, then the other nodes become simplicial. Thus, such sets can be ordered together. With \emph{path compression} we replace any path of nodes with degree 2 by a single degree-2 node. If one node on the path is eliminated, then its degree-2 neighbors can be eliminated next in a minimum fill-in ordering.

\new{\emph{Degree-2 elimination} is an inexact reduction rules that eliminates nodes of degree 2.}
This reduction turns out to be exact if none of the eliminated nodes are also separators. Lastly, \emph{triangle contraction} contracts adjacent nodes of degree 3 that share at least one neighbor.

To our knowledge, only the indistinguishable node reduction has been used in practice. While linear time algorithms for ordering chordal graphs are known, it appears that the special structure of simplicial nodes is not exploited in non-chordal graphs.
We now describe the reduction rules in greater detail. \ifTR\else Proofs of the statements can be found in Appendix \ref{sec:proofs}.\fi{}\\

% Reduction rules
%%%%%%%%%%%%%%%%%%%%%%%%%%%%%%%%%%%%%%%%%
\begin{figure}[b!]
        \centering
        \small
        \begin{tikzpicture}[x=0.9cm, y=1.0cm]

%%%%%%%%%%%%%%%%%%%%
% Simplicial Nodes %
%%%%%%%%%%%%%%%%%%%%

\begin{scope}[shift={(-3.75, 0)}]
\node (simplicial) at (1, 2.0) {Simplicial Nodes};
\begin{scope}[every node/.style={circle, draw}]
        \node[thick] (A) at (0,  0.5) {$s$};
        \node        (B) at (1,  1.0) {};
        \node        (C) at (1,  0.0) {};
        \node        (D) at (2,  1.0) {};
        \node        (E) at (2,  0.0) {};
\end{scope}

\begin{scope}
        \draw (A) to (B);
        \draw (A) to (C);
        \draw (A) to (D);
        \draw (A) to (E);
        \draw (B) to (C);
        \draw (B) to (D);
        \draw (B) to (E);
        \draw (C) to (D);
        \draw (C) to (E);
        \draw (D) to (E);

        \draw[dashed] (B) -- +(0,  0.5);
        \draw[dashed] (C) -- +(0, -0.5);
        \draw[dashed] (D) -- +(0,  0.5);
        \draw[dashed] (E) -- +(0, -0.5);
\end{scope}
\end{scope}

%% Separator
\draw (-1.1, -0.5) -- (-1.1, 2.25);
\hspace*{.5cm}
%%%%%%%%%%%%%%%%%%%%%%%%%%%
% Indistinguishable Nodes %
%%%%%%%%%%%%%%%%%%%%%%%%%%%

\node (indistinguishable) at (0.5, 2.0) {Indistinguishable Nodes};

\begin{scope}[every node/.style={circle, draw}]
        \node[thick] (A) at ( 0.00,  0) {$i_1$};
        \node[thick] (B) at ( 1.00,  0) {$i_2$};
        \node        (C) at (-0.25,  1) {};
        \node        (D) at ( 0.50,  1) {};
        \node        (E) at ( 1.25,  1) {};
\end{scope}

\begin{scope}
        \draw[thick] (A) to (B);
        \draw (A) to (C);
        \draw (A) to (D);
        \draw (A) to (E);
        \draw (B) to (C);
        \draw (B) to (D);
        \draw (B) to (E);

        \draw (C) -- (D);

        \draw[dashed] (C) -- +( 0.00, 0.5);
        \draw[dashed] (C) -- +(-0.25, 0.5);
        \draw[dashed] (C) -- +( 0.25, 0.5);
        \draw[dashed] (D) -- +( 0.00, 0.5);
        \draw[dashed] (E) -- +( 0.00, 0.5);
        \draw[dashed] (E) -- +( 0.25, 0.5);
\end{scope}

%% Separator
\hspace*{.5cm}
\draw (2.1, -0.5) -- (2.1, 2.25);

%%%%%%%%%
% Twins %
%%%%%%%%%

\node (twins) at (3.5, 2.0) {Twins};

\begin{scope}[every node/.style={circle, draw}]
        \node[thick] (A) at ( 3.00,  0) {$t_1$};
        \node[thick] (B) at ( 4.00,  0) {$t_2$};
        \node        (C) at ( 2.75,  1) {};
        \node        (D) at ( 3.50,  1) {};
        \node        (E) at ( 4.25,  1) {};
\end{scope}

\begin{scope}
        \draw (A) to (C);
        \draw (A) to (D);
        \draw (A) to (E);
        \draw (B) to (C);
        \draw (B) to (D);
        \draw (B) to (E);

        \draw (C) -- (D);

        \draw[dashed] (C) -- +( 0.00, 0.5);
        \draw[dashed] (C) -- +(-0.25, 0.5);
        \draw[dashed] (C) -- +( 0.25, 0.5);
        \draw[dashed] (D) -- +( 0.00, 0.5);
        \draw[dashed] (E) -- +( 0.00, 0.5);
        \draw[dashed] (E) -- +( 0.25, 0.5);
\end{scope}

\end{tikzpicture}
        \caption{Examples for simplicial nodes, indistinguishable nodes and twins. The neighborhood of $s$ is a clique, so $s$ is simplicial. Nodes $i_1$ and $i_2$ are indistinguishable, since they are neighbors and adjacent to all unlabeled nodes, i.e., $\Gamma[i_1] = \Gamma[i_2]$. Nodes $t_1$ and $t_2$ are twins, since they are both adjacent to all unlabeled nodes, but not to each other. $\Gamma(t_1) = \Gamma(t_2)$.}
        \label{fig:indistinguishable-twins}
\end{figure}
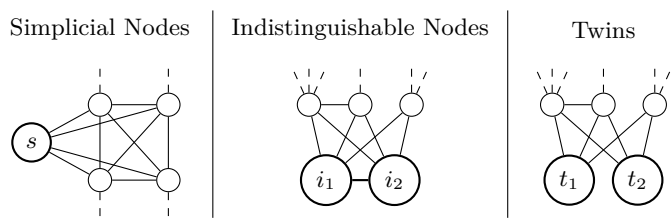
\vspace*{-.5cm}
\subsection{The Simplicial Node Reduction}
A node $x$ is \emph{simplicial} if its neighborhood $\Gamma(x)$ is a clique (see Figure \ref{fig:indistinguishable-twins} for an example). There exists a minimum fill-in ordering where $x$ is eliminated first.
\begin{theorem}
        \label{thm:simplicial}
        Let $G = (V, E)$ be a graph with a simplicial node $x$. The ordering $x \Sigma(G_x)$ is a minimum fill-in ordering of $G$.
\end{theorem}
\ifTR
\begin{proof}
        Since $\Gamma(x)$ is a clique, $D(x) = \emptyset$. The fill-in associated with eliminating $x$ first is $\phi(G, x\Sigma(G_x)) = |D(x)| + \Phi(G_x) = \Phi(G_x)$. From \eqref{eq:phi-non-decreasing} it follows that $\phi(G, x\Sigma(G_x)) = \Phi(G)$.
\end{proof}\fi{}
This allows us to eliminate all simplicial nodes first by the following procedure: Find any simplicial node $x$ in $G = (V, E)$,
         eliminate $x$ from $G$ and place it next in the node ordering.
         If the elimination graph $G_x$ has simplicial nodes, then repeat the procedure for $G_x$.
If every elimination graph in the elimination sequence $\sigma$ has at least one simplicial node, then $\phi(G,\sigma) = 0$. In this case, $\sigma$ is a perfect elimination ordering of $G$. Graphs that admit such an ordering are called \emph{chordal} or \emph{triangulated} graphs~\cite{rose70,rose72}.
\begin{reduction}[Simplicial Node Reduction]
        \label{red:simplicial}
        Given a graph $G = (V, E)$ and a simplicial node $x \in V$, construct a new graph $G' = G[V \setminus \{x\}]$. $\Phi(G) = \Phi(G')$ and $x \Sigma(G')$ is a minimum fill-in ordering of $G$.
\end{reduction}

%%%%%%%%%%%%%%%%%%%%%%%%%%%%%%%%%%%%%%%%%%%%%%%%
\subsection{The Indistinguishable Node Reduction}
\label{sec:indistinguishable}
Two nodes $a$ and $b$ are \emph{indistinguishable} if $\Gamma[a] = \Gamma[b]$ (see Figure \ref{fig:indistinguishable-twins} for an example). Such nodes can be eliminated together: if $a$ and $b$ are indistinguishable nodes, then there exists a minimum fill-in ordering $x_1 \cdots x_i a b x_{i+1} \cdots x_\ell$, where $\{x_1,\dots,x_i,x_{i+1},\dots,x_\ell\} = V \setminus \{a, b\}$. To obtain a reduced graph $G'$, we contract a set of indistinguishable nodes $S$ in $G$ to a single node.

We first establish that indistinguishable nodes stay indistinguishable throughout the elimination sequence. Then, we show that eliminating indistinguishable nodes as described does in fact lead to minimum fill-in orderings.

\begin{lemma}
        \label{lem:indistinguishable-1}
        If $a$, $b$ are indistinguishable nodes in a graph $G$, then $a$ and $b$ are indistinguishable in any elimination graph $G_x$ for $x \notin \{a, b\}$.
\end{lemma}
\ifTR
\begin{proof}
        Let $x \in \Gamma(a) \setminus \{b\} = \Gamma(b) \setminus \{a\}$ be eliminated from $G$.
        In the elimination graph $\Gamma_{G_x}(a) = (\Gamma(a) \setminus \{x\}) \cup \Gamma(x)$ and $\Gamma_{G_x}(b) = (\Gamma(b) \setminus \{x\}) \cup \Gamma(x)$. Since $a \in \Gamma_{G_x}(b)$ and $b \in \Gamma_{G_x}(a)$, $\Gamma_{G_x}[a] = \Gamma_{G_x}[b]$. Thus, $a$ and $b$ are indistinguishable in $G_x$.
        
        If a node $y$ with $y \notin \Gamma(a)$ and $y \notin \Gamma(b)$ is eliminated from $G$, the neighborhoods of $a$ and $b$ do not change, since $a, b \notin \Gamma(y)$. In the elimination graph $\Gamma_{G_y}[a] = \Gamma_{G_y}[b]$.
        Thus, $a$ and $b$ are indistinguishable in $G_y$.
\end{proof}\fi{}

\begin{theorem}
        \label{thm:indistinguishable}
        Let $G = (V, E)$ be a graph with a set of nodes $A \subseteq V$, where $\forall\ a_i, a_j \in A,\ \Gamma[a_i] = \Gamma[a_j]$. There is an ordering $\sigma' = x_1 \cdots x_i A x_{i+1} \cdots x_\ell$, where $V \setminus A = \{x_1,\dotsc,x_\ell\}$, such that $\phi(G, \sigma') = \Phi(G)$.
\end{theorem}
\ifTR\begin{proof}
        Lemma \ref{lem:indistinguishable-1} implies that all pairs of nodes in $A$ are indistinguishable in all graphs in the elimination sequence.
        Let $a \in A$ be the node that is eliminated before all other nodes in $A$.
        There is a graph $G^{(m)}$ in the elimination sequence with a minimum ordering $a \Sigma(G^{(m)}_a),\ a \in A$.
        For all $b \in A \setminus \{a\}$ $\Gamma_{G^{(m)}_a}(b)$ is a clique, i.e., these nodes are simplicial after elimination of $a$.
        Thus, $A \Sigma(G^{(m)}_A)$ is a minimum ordering of $G^{(m)}$ and G has a minimum ordering of the form of $\sigma'$.
\end{proof}\fi{}

\begin{reduction}[Indistinguishable Node Reduction]
        \label{red:indistinguishable}
        Given a graph $G = (V, E)$ with indistinguishable nodes $a, b \in V$, construct a new graph $G' = G(V \setminus \{b\})$. Replacing $a$ in $\Sigma(G')$ by $ab$ results in a minimum ordering of $G$.
\end{reduction}
Note, that in the reduced graph $G'$, the deficiency of any node neighboring a set of indistinguishable nodes is different from that of the corresponding node in the original graph $G$. Thus, we have to optimize the ordering in $G'$ not in terms of the deficiency of a node in $G'$, but in terms of the deficiency of the corresponding node in $G$. 

Indistinguishable nodes are commonly used to speed up the minimum degree algorithm~\cite{george78,george80,george1989md}. In this context the reduction has been shown to be exact. This reduction is also known as \emph{graph compression} and is used in other variants of nested dissection and the minimum degree algorithm, see for example the algorithms by Ashcraft~\cite{ashcraft95} and Hendrickson and Rothberg~\cite{hendrickson98}. \\

%%%%%%%%%%%%%%%%%%%%%%%%%%%%%%
        \vspace*{-.5cm}
\subsection{The Twin Reduction}
Two nodes $a$ and $b$ are \emph{twins} if $\Gamma(a) = \Gamma(b)$ (see Figure \ref{fig:indistinguishable-twins} for an example). Similar to indistinguishable nodes, twins can be eliminated together.
\begin{theorem}
        Let $a$, $b$ be twins in a graph $G = (V, E)$. There exists an ordering $\sigma' = x_1 \cdots x_i a b x_{i+1} \cdots x_l$, with $x_j \in V \setminus \{a, b\}$, such that $\phi(G, \sigma') = \Phi(G)$.
\end{theorem}
\ifTR\begin{proof}
        If a node $x \in \Gamma(a) = \Gamma(b)$, is eliminated, $a$ and $b$ form a clique in the elimination graph $G_x$. Thus, $a$ and $b$ are indistinguishable in $G_x$ and Theorem \ref{thm:indistinguishable} holds.
        If a node $x \notin \Gamma(a) \cup \{a, b\}$ is eliminated, the neighborhoods of nodes $a$ and $b$ do not change, i.e., $\Gamma_{G_x}[a] = \Gamma_G[a]$ and $\Gamma_{G_x}[b] = \Gamma_G[b]$. Thus, $a$ and $b$ are twins in $G_x$.
        If $a$ is eliminated, $\Gamma_{G_a}(b)$ is a clique in the elimination graph $G_a$ and $b$ is simplicial in $G_a$. With Theorem \ref{thm:simplicial}, $b \Sigma((G_a)_b)$ is a minimum ordering of $G_a$ and $a b \Sigma((G_a)_b)$ is a minimum ordering of $G$.
\end{proof}\fi{}
We can treat twins similarly to indistinguishable nodes: we obtain a reduced graph by contracting twins. As with Reduction \ref{red:indistinguishable}, the deficiency of a node neighboring contracted twins in $G'$ is smaller than the deficiency of the corresponding node in $G$. Thus, orderings of $G'$ should be evaluated not in terms of the deficiency of nodes in $G'$, but in terms of the deficiency of corresponding nodes in $G$.
\begin{reduction}[Twin Reduction]
        \label{red:twin}
        Given a graph $G = (V, E)$ with twins $a, b \in V$, construct a new graph $G' = G[V \setminus \{b\}]$. Replacing $a$ in $\Sigma(G')$ by $ab$ results in a minimum ordering of $G$.
\end{reduction}

%%%%%%%%%%%%%%%%%%%%%%%%%%%%%%%%
\subsection{Path Compression}
We now show that a path of nodes with degree 2 can be eliminated together. More formally, let $P = \{a_1, a_2, \dotsc, a_k\}$ be a path in a graph $G = (V, E)$ with $\deg(a_i) = 2$ for all $a_i \in P$. There is a minimum fill-in ordering $\Sigma = x_1 \cdots x_i a_1 \cdots a_k x_{i+1} \cdots x_\ell$, where $V \setminus P = \{x_1,\dotsc,x_\ell\}$.

We prove this by distinguishing three cases based on which nodes are separation cliques, and using the relationship between minimum triangulations and minimum fill-in orderings. Corollary 1 and Proposition 2 from~\cite{rose72} are central to our proof and we restate them here.
\begin{lemma}[Corollary 1 from~\cite{rose72}]
        \label{lem:sep-clique}
        Let $G = (V, E)$ be a graph with separation clique $S$ with components $C_1, C_2, \dotsc, C_k$. Any minimum triangulation $T$ of $G$ contains only edges $e = \{x, y\} \in T$ with $x$ and $y$ in the same component $C_j$, or edges $e = \{x, y\} \in T$ with $x \in C_j$ and $y \in S$.
\end{lemma}
\begin{lemma}[Proposition 2 from~\cite{rose72}]
        \label{lem:cycle}
        Let $C = (V, E)$ be a cycle with $|V| \geq 3$ nodes. Any ordering of $C$ is a minimum fill-in ordering.
\end{lemma}
Furthermore, we need to show that nodes with degree 2 in induced cycles of four or more nodes can be eliminated first.
\begin{lemma}
        \label{lem:deg-2-nodes}
        Let $G = (V,E)$ be a graph with a node $a \in V$ where $\deg(a) = 2$, $\Gamma(a) \notin E$ and $\{a\}$ is not a separation clique. Then, $a\Sigma(G_a)$ is a minimum ordering of $G$.
\end{lemma}
To prove Lemma \ref{lem:deg-2-nodes} we establish that there exists a minimum triangulation that does not contain an edge to such a node $a$.
\begin{lemma}
        \label{lem:deg-2-triangulation}
        Let $G$ and $a$ be as in Lemma \ref{lem:deg-2-nodes}. There exists a minimum triangulation $\hat{T}$ of $G$, with $\Gamma(a) \in \hat{T}$ and $\{a, x\} \notin \hat{T}$ for all $x \in V$.
\end{lemma}
\ifTR\begin{proof}
        Let $\mathcal{C} = \{C_1,\dotsc,C_n\}$ be the set of induced cycles that contain $a$, i.e., for all $i$, $a \in C_i$ and $G[C_i]$ is a cycle. Since $a$ has degree two, $\Gamma(a) \subset C_i$ for all $i$.
        By Lemma \ref{lem:cycle}, for all $C_i \in \mathcal{C}$, there exists a minimum triangulation $T_i$ of $G[C_i]$ with $\Gamma(a) \in T_i$. Thus, there exists a minimum triangulation $\hat{T}$ of $G$ with $\Gamma(a) \in \hat{T}$.
        $\Gamma(a)$ is a separation clique with components $\{a\}$ and $V \setminus (\{a\} \cup \Gamma(a))$ in the triangulated graph $\hat{G} = (V, E \cup \hat{T})$. By Lemma \ref{lem:sep-clique} there exists no edge $\{a, x\} \in \hat{T}$.
        This implies $\Gamma(a) \in \hat{T}$, $\{a, x\} \notin \hat{T}$ and $\hat{T}$ is minimum.
\end{proof}
\begin{proof}[Proof of Lemma \ref{lem:deg-2-nodes}]
        With Lemma \ref{lem:deg-2-triangulation} there exists a minimum triangulation $\hat{T}$ of $G$ with $\Gamma(a) \in \hat{T}$ and $\{a, x\} \notin \hat{T}$. $a$ is simplicial in the triangulated graph $\hat{G} = (V, E \cup \hat{T})$ and $a\Sigma(\hat{G}_a)$ is a minimum ordering of $\hat{G}$. This implies that $a \Sigma(G_a)$ is a minimum ordering of $G$. Note that eliminating $a$ from $G$ adds the edge $\Gamma(a)$ to the elimination graph.
\end{proof}
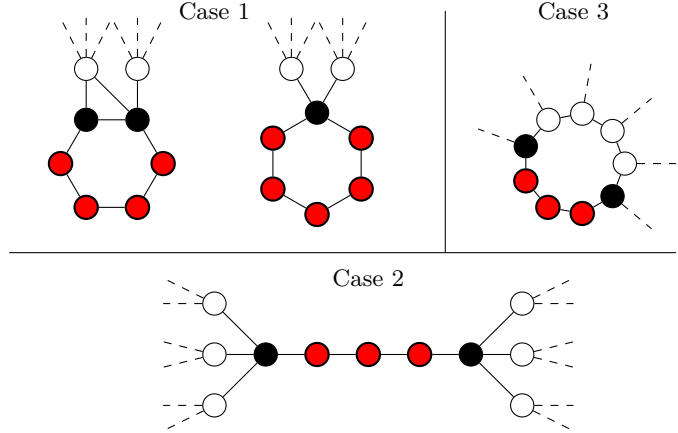
\begin{figure}
        \centering
        \small
        \begin{tikzpicture}[x=0.75cm,y=0.75cm, scale=0.9]

%%%% Case 1: neighborhood is a separator

\begin{scope}
        \node (Case1) at (2, 3) {Case 1};

        % Example with two connected neighbors
        \begin{scope}[every node/.style={circle, draw}]
                \node[thick, fill=red] (A1) at (-1.00,  0.00) {}; % {$a_1$};
                %% N(P)
                \node[fill=black] (B1) at (-0.50,  0.86) {}; % {$a_0$};
                \node[fill=black] (C1) at ( 0.50,  0.86) {}; % {$a_5$};
                %% end of N(P)
                \node[thick, fill=red] (D1) at ( 1.00,  0.00) {}; % {$a_4$};
                \node[thick, fill=red] (E1) at ( 0.50, -0.86) {}; % {$a_3$};
                \node[thick, fill=red] (F1) at (-0.50, -0.86) {}; % {$a_2$};

                \node (X1) at (-0.50, 1.86) {};
                \node (Y1) at ( 0.50, 1.86) {};
        \end{scope}

        \begin{scope}
                \draw (A1) -- (B1);
                \draw (B1) -- (C1);
                \draw (C1) -- (D1);
                \draw (D1) -- (E1);
                \draw (E1) -- (F1);
                \draw (F1) -- (A1);

                \draw (B1) -- (X1);
                \draw (C1) -- (X1);
                \draw (C1) -- (Y1);

                \draw[dashed] (X1) -- +(-0.5, 1);
                \draw[dashed] (X1) -- +( 0.0, 1);
                \draw[dashed] (X1) -- +( 0.5, 1);
                \draw[dashed] (Y1) -- +(-0.5, 1);
                \draw[dashed] (Y1) -- +( 0.0, 1);
                \draw[dashed] (Y1) -- +( 0.5, 1);
        \end{scope}

        % Example with one neighbor
        \begin{scope}[shift={(4, 0)}]
                \begin{scope}[every node/.style={circle, draw}]
                        \node[thick, fill=red] (A2) at ( 0.00, -1.00) {}; % {$a_3$};
                        \node[thick, fill=red] (B2) at (-0.86, -0.50) {}; % {$a_2$};
                        \node[thick, fill=red] (C2) at (-0.86,  0.50) {}; % {$a_1$};
                        \node[fill=black]      (D2) at ( 0.00,  1.00) {}; % {$a_0$};
                        \node[thick, fill=red] (E2) at ( 0.86,  0.50) {}; % {$a_5$};
                        \node[thick, fill=red] (F2) at ( 0.86, -0.50) {}; % {$a_4$};

                        \node (X2) at (-0.50, 1.86) {};
                        \node (Y2) at ( 0.50, 1.86) {};
                \end{scope}

                \begin{scope}
                        \draw (A2) -- (B2);
                        \draw (B2) -- (C2);
                        \draw (C2) -- (D2);
                        \draw (D2) -- (E2);
                        \draw (E2) -- (F2);
                        \draw (F2) -- (A2);

                        \draw (D2) -- (X2);
                        \draw (D2) -- (Y2);

                        \draw[dashed] (X2) -- +(-0.5, 1);
                        \draw[dashed] (X2) -- +( 0.0, 1);
                        \draw[dashed] (X2) -- +( 0.5, 1);
                        \draw[dashed] (Y2) -- +(-0.5, 1);
                        \draw[dashed] (Y2) -- +( 0.0, 1);
                        \draw[dashed] (Y2) -- +( 0.5, 1);
                \end{scope}
        \end{scope}
        %%% Separator line
        \draw (-2, -1.75) -- (11, -1.75);
\end{scope}

%%%% Case 2: nodes in P are separators

\begin{scope}[shift={(3.0, -3.75)}]
        \node (Case2) at (2, 1.5) {Case 2};

        \begin{scope}
                \begin{scope}[every node/.style={circle, draw}]
                        \node[fill=black]      (A) at (0, 0) {}; % {$a_0$};
                        \node[thick, fill=red] (B) at (1, 0) {}; % {$a_1$};
                        \node[thick, fill=red] (C) at (2, 0) {}; % {$a_2$};
                        \node[thick, fill=red] (D) at (3, 0) {}; % {$a_3$};
                        \node[fill=black]      (E) at (4, 0) {}; % {$a_4$};

                        \node (X1) at (-1,  1) {};
                        \node (X2) at (-1,  0) {};
                        \node (X3) at (-1, -1) {};

                        \node (Y1) at (5,  1) {};
                        \node (Y2) at (5,  0) {};
                        \node (Y3) at (5, -1) {};
                \end{scope}

                \begin{scope}
                        \draw (A) -- (B);
                        \draw (B) -- (C);
                        \draw (C) -- (D);
                        \draw (D) -- (E);

                        \draw (A) -- (X1);
                        \draw (A) -- (X2);
                        \draw (A) -- (X3);

                        \draw (E) -- (Y1);
                        \draw (E) -- (Y2);
                        \draw (E) -- (Y3);

                        \draw[dashed] (X3) -- +(-1, -0.50);
                        \draw[dashed] (X3) -- +(-1,  0.00);
                        \draw[dashed] (X2) -- +(-1, -0.25);
                        \draw[dashed] (X2) -- +(-1,  0.25);
                        \draw[dashed] (X1) -- +(-1,  0.50);
                        \draw[dashed] (X1) -- +(-1,  0.00);
                        \draw[dashed] (Y3) -- +( 1, -0.50);
                        \draw[dashed] (Y3) -- +( 1,  0.00);
                        \draw[dashed] (Y2) -- +( 1, -0.25);
                        \draw[dashed] (Y2) -- +( 1,  0.25);
                        \draw[dashed] (Y1) -- +( 1,  0.50);
                        \draw[dashed] (Y1) -- +( 1,  0.00);
                \end{scope}
        \end{scope}
\end{scope}

%%%% Case 3: neither nodes in P, nor N(P) are separators

\begin{scope}[shift={(9, 0)}] 
\node (Case3) at (0, 3) {Case 3};

\begin{scope}[every node/.style={circle, draw}]
        \node        (A) at ( 1.00,  0.00) {};
        \node        (B) at ( 0.76,  0.64) {};
        \node        (C) at ( 0.17,  0.98) {};
        \node        (D) at (-0.49,  0.86) {};
        \node[fill=black]      (E) at (-0.93,  0.34) {}; % {$a_0$};
        \node[thick, fill=red] (F) at (-0.93, -0.34) {}; % {$a_1$};
        \node[thick, fill=red] (G) at (-0.50, -0.86) {}; % {$a_2$};
        \node[thick, fill=red] (H) at ( 0.17, -0.98) {}; % {$a_3$};
        \node[fill=black]      (I) at ( 0.76, -0.64) {}; % {$a_4$};
\end{scope}

\begin{scope}
        \draw (A) -- (B);
        \draw (B) -- (C);
        \draw (C) -- (D);
        \draw (D) -- (E);
        \draw (E) -- (F);
        \draw (F) -- (G);
        \draw (G) -- (H);
        \draw (H) -- (I);
        \draw (I) -- (A);

        \draw[dashed] (A) -- +( 1.00,  0.00);
        \draw[dashed] (B) -- +( 0.76,  0.64);
        \draw[dashed] (C) -- +( 0.17,  0.98);
        \draw[dashed] (D) -- +(-0.49,  0.86);
        \draw[dashed] (E) -- +(-0.93,  0.34);
        \draw[dashed] (I) -- +( 0.76, -0.64);
\end{scope}
        %%% Separator line
        \draw (-2.5, -1.75) -- (-2.5, 3);
\end{scope}

\end{tikzpicture}
        \caption{Examples for the three cases in the proof of Theorem \ref{thm:pathcompression}. Red nodes are nodes in $P$, black nodes are in $\Gamma(P)$. Dashed edges lead to some other nodes in the graph.}
        \label{fig:path-cases}
\end{figure}\fi{}
With these results we can now prove our original statement.
\begin{theorem}
        \label{thm:pathcompression}
        Let $G = (V, E)$ and $P = \{a_1,\dotsc, a_k\} \subseteq V$ such that $G[P]$ is a path graph and $\forall\ a \in P\,\,\deg(a) = 2$. Let $\Gamma(P) = \{a_0, a_{k+1} \}$ and $\Gamma(a_i) = \{a_{i-1}, a_{i+1}\}$, $i = 1,\dotsc,k$. There exists an ordering $\sigma' = x_1 \cdots x_i a_1 \cdots a_k x_{i+1} \cdots x_\ell$ where $V \setminus P = \{x_1,\dotsc,x_\ell\}$, such that $\phi (G, \sigma') = \Phi(G)$.
\end{theorem}\ifTR
\begin{proof}
        $G$ can be decomposed into non-disjoint graphs $G' := G[V\setminus P]$ and $G'' := G[P \cup \Gamma(P)]$, such that $G = G' \cup G''$.
        We distinguish three cases (see Figure \ref{fig:path-cases} for examples):
        \begin{description}
                \item[Case 1:] If $a_0 = a_{k+1}$ or $a_0 \in \Gamma(a_{k+1})$, then
                        $G''$ is a cycle and $\Gamma(P)$ is a separation clique with leaves $G'$ and $G''$. Let $T'$ be a minimum triangulation of $G'$ and $T''$ be a minimum triangulation of $G''$. By Lemma \ref{lem:sep-clique}, $T' \cup T''$ is a minimum triangulation of $G$. Since any ordering of $G''$ generates a minimum triangulation of $G''$ (by Lemma \ref{lem:cycle}), $P\Sigma(G''_P)$ is a minimum ordering of $G''$ and $P\Sigma(G_P)$ is a minimum ordering of $G$.
                \item[Case 2:] If $a_0 \neq a_{k+1}$, and $\{a_0\}$ and $\{a_{k+1}\}$ are separation cliques,
                        then all nodes in $P$ are also separation cliques. By Lemma \ref{lem:sep-clique}, there are no edges $\{a_i, a_j\}$, for all $i \neq j$ in a minimum triangulation of $G$.

                        Let $\Sigma$ be any minimum fill-in ordering of $G$ and let $G^{(m)}$ be the graph in the elimination sequence from which $a \in P$ is eliminated. Node $a$ is simplicial in $G^{(m)}$, otherwise $T(\Sigma)$ would not be a minimum triangulation. Since all $a \in P$ are separation cliques and $\deg(a) = 2$ in $G$, $\deg(a) = 1$ in $G^{(m)}$.

                        Without loss of generality assume that $a_1$ is eliminated before all other nodes in $P$. Let $G^{(m_1)}$ be the graph in the elimination sequence from which $a_1$ is eliminated. If $\deg(a_1) = 1$ in $G^{(m_1)}$, then $\deg(a_2) = 1$ in $G^{(m_1)}_{a_1}$. Repeating this argument for all $a_i \in P$ proves that $P\Sigma(G^{(m_1)}_P)$ is a minimum ordering of $G^{(m_1)}$ and $\Sigma$ is of the form of $\sigma'$.
                \item[Case 3:] If $\{a_0\}$, $\{a_{k+1}\}$ and $\Gamma(P)$ are not separation cliques, then any $a \in P$ satisfies the conditions in Lemma \ref{lem:deg-2-nodes}. In $G_{a}$, $\{a_0\}$, $\{a_{k+1}\}$ and $\Gamma(P)$ are not separation cliques. Repeating the argument for $G_a$ leads to a minimum ordering $P \Sigma(G_P)$.
        \end{description}
        In Case 1 and Case 3, there exists a minimum ordering $a_1 \cdots a_k x_1 \cdots x_{\ell}$. In Case 2, there exists a minimum ordering $x_1 \cdots x_i a_1 \cdots a_k x_{i+1} \cdots x_\ell$. Both orderings are of the form of~$\sigma'$.
\end{proof}\fi{}
Since such sets of nodes $P$ can be eliminated together, we can contract them to a single node. It is possible that in a minimum elimination sequence of a graph $G$, the degree of $a_1 \in P$ becomes 1. Then, $P$ has to be ordered as $a_1a_2\cdots a_k$ to obtain a minimum ordering.
\begin{reduction}[Path Compression]
        \label{red:path}
        Given a graph $G = (V, E)$ with a set of nodes $P = \{a_1,\dotsc,a_k\}$, where $G[P]$ is a path graph, $N(P) = \{a_0, a_{k+1}\}$ and $\forall\ a \in P\ \deg(a) = 2$, construct a new graph $G' = (V \setminus \{a_2,\dotsc,a_k\}, E')$, where $E' = \left(E \setminus E(P \cup \{a_{k+1}\})\right) \cup \{\{a_1, a_{k+1}\}\}$. Replacing $a_1$ in $\Sigma(G')$ by $a_1a_2\cdots a_k$ yields a minimum ordering of $G$.
\end{reduction}

%%%%%%%%%%%%%%%%%%%%%%%%%%%%%%%%%%%%
\subsection{Degree-2 Elimination} 
\label{sec:deg-2}

\begin{inexactreduction}[Degree-2 Elimination]
        \label{red:degree-2}
        Given a graph $G = (V, E)$ and any node $x$ with degree 2, construct the elimination graph $G_x$. The potentially non-minimum ordering of $G$ is $x \Sigma(G_x)$. The reduction is applied recursively until no nodes with degree 2 are left.
\end{inexactreduction}
The proof of Theorem \ref{thm:pathcompression} has interesting implications on the exactness of degree-2 elimination.
\begin{corollary}
        \label{cor:cycle-deg2}
        Let $G = (V, E)$ be a graph. If $x \in V$ is in any cycle $C \subseteq V$ and $\deg(x) = 2$, $x\Sigma(G_x)$ is a minimum ordering of $G$.
\end{corollary}
\ifTR\begin{proof}
        Node $x$ is part of a cycle and thus not a separation clique. Either case 1 or 3 of Theorem \ref{thm:pathcompression} holds, which implies that $x \Sigma(G_x)$ is a minimum ordering of $G$.
\end{proof}\fi{}
\begin{corollary}
        \label{cor:sep-deg2}
        Let $G = (V, E)$ be a graph. Let $x \in V$ be a separation clique and $\deg(x) = 2$.
        Let $\Sigma$ be a minimum fill-in ordering and $G^{(i)}$ be the graph in the corresponding elimination sequence,
        from which $x$ is eliminated, i.e., $\Sigma(i+1) = x$.
        The node $x$ is simplicial in $G^{(i)}$.
\end{corollary}
\ifTR\begin{proof}
        Since $x$ is a separation clique, Case 2 of Theorem~\ref{thm:pathcompression} holds and thus, $x$ is simplicial in~$G^{(i)}$.
\end{proof}\fi{}
Corollaries \ref{cor:cycle-deg2} and \ref{cor:sep-deg2} imply that degree-2 elimination is exact if only degree-2 nodes that are part of a cycle are eliminated. In graphs where no degree-2 nodes are separators, degree-2 elimination is therefore exact.\\

%%%%%%%%%%%%%%%%%%%%%%%%%%%%%%%%%%%%

\vspace*{-.5cm}
\subsection{Triangle Contraction}
\label{sec:triangle}
%\noindent\textbf{Triangle Contraction.}
Consider two adjacent nodes $a, b \in V$ where $\deg(a) = \deg(b) = 3$ and $|\Gamma(a) \cap \Gamma(b)| \geq 1$. Eliminating node $a$ does not increase the degree of node $b$, and vice versa. Note, that after eliminating $a$, $|D(b)| \leq 2$, i.e., eliminating $b$ only inserts two edges into the graph. Since this fill-in is small, we eliminate $b$ as soon as $a$ was eliminated, and vice versa. Thus, we contract nodes $a$ and $b$.

\begin{inexactreduction}[Triangle Contraction]
        \label{red:triangle}
        Given a graph $G = (V, E)$ and nodes $a, b$ with $\deg(a) = \deg(b) = 3$ and $|\Gamma(a) \cap \Gamma(b)| = 1$, construct a new graph $G' = (V \setminus \{a\}, E \setminus (\cup_{x \in \Gamma(a)} \{a, x\}) \cup_{x \in \Gamma(a)} \{x, b\})$. Replacing $b$ by $ba$ in $\Sigma(G')$ yields a potentially non-minimum ordering of $G$.
\end{inexactreduction}

\ifTR
\section{Implementation Details}
\label{sec:implementation}
To apply simplicial node reduction (Reduction \ref{red:simplicial}), we iterate through nodes in order by non-decreasing degree. To test if a node $x$ is simplicial, we iterate through the neighbors $y \in \Gamma(x)$. If $|\Gamma(y) \cap \Gamma(x)| = \deg(x) - 1$ for all $y$, then $x$ is simplicial. When a node is found to be simplicial, we mark it as removed and adjust the degrees of its neighbors accordingly. Removed nodes are ignored when testing the other nodes. The order in which simplicial nodes are found yields their elimination order. Since we only evaluate each node once in a single pass, this method may introduce new simplicial nodes that remain in the graph. However, in practice we find that most simplicial nodes are eliminated in a single pass.

Deciding if a node $v$ is simplicial takes time $O(\deg(v)^2)$.
For graphs where $\deg(v) = O(n)$ this implies a total time for simplicial node reduction of $O(n^3)$.
To avoid this case, we introduce a parameter $\Delta$ and only test nodes $v$ that have degree $\deg(v) \leq \Delta$.
The total time for simplicial node reduction is then $O(n \Delta^2)$.

The indistinguishable node and twin reductions (Reductions \ref{red:indistinguishable} and \ref{red:twin}) are similar in their implementation and are based on the algorithms by Ashcraft \cite{ashcraft95} and Hendrickson and Rothberg \cite{hendrickson98}. For both reductions we first compute a hash of the neighborhood of each node $x_i$ as
$h_c(x_i) = \sum_{y_j \in \Gamma[x_i]} j$ and
$h_o(x_i) = \sum_{y_j \in \Gamma(x_i)} j$
We only compare the neighborhoods directly if the hashes of two candidates are equal.
To detect indistinguishable nodes, we now go through all pairs $(u, v)$ of adjacent nodes and, if $h_c(u) = h_c(v)$,
test if $\Gamma[u] = \Gamma[v]$. Detecting and contracting sets of indistinguishable nodes in this way takes time $O(m)$.
To detect twins, we first sort the list of hashes $h_o$. We then go through the list, and, for pairs of nodes $(u,v)$ with equal hash and degree, test if $\Gamma(u) = \Gamma(v)$.
In the worst case, if all hashes are equal and all nodes have the same degree, our implementation takes time $O(mn + n \log(n))$.

In path compression (Reduction \ref{red:path}) and degree-2 elimination (Reduction \ref{red:degree-2}), nodes to contract or eliminate are detected in time $O(n)$. The reduced graph is then built in time $O(m)$.
We order sets of nodes contracted by to path compression starting at the end whose neighbor is eliminated first.
Nodes removed during degree-2 elimination appear in the final ordering as they are removed from the graph.  

We detect set of nodes $A$ to be contracted in triangle contraction (Reduction \ref{red:triangle}) by the following procedure:
         Let $x$ be some node with $\deg(x) = 3$. Add $x$ to $A$.
         Then we repeat the following procedure: If $x$ has a neighbor $y$ with $\deg(y) = 3$ and $|\Gamma(x) \cap \Gamma(y)| \geq 1$, add $x$ and $y$ to $A$. Let $a \in (\Gamma(x) \cap \Gamma(y))$.
         Let $z \in \Gamma(y), z \notin A$. If $\deg(z) = 3$ and $a \in \Gamma(z)$, add $z$ to $A$. Otherwise, stop.
        Repeat the procedure with the neighbors of $z$.
This reduction can be implemented in time $O(m)$.
In the ordering of the input graph, nodes in $A$ are ordered as they are added to $A$.\fi

\section{Experimental Evaluation}
\label{s:experiments}
\textbf{Methodology.}
\ifTR We implemented the reductions in \mbox{C++} within version 2.10 of the KaHIP graph partitioning framework \cite{kaHIPHomePage} and compiled using \mbox{g++ 8.3.0} with optimization flag \texttt{-O3}. 
\else We implemented the reductions in \mbox{C++}  and compiled using \mbox{g++ 8.3.0} with optimization flag \texttt{-O3}. Additional implementation details can be found in Appendix~\ref{s:impdetails}.\fi{}
We use Metis (version 5.0) \cite{karypis1998fast} to perform nested dissection. All running times were measured on a machine with four Intel Xeon E7-8867 v3 processors (16 cores, 2.5 GHz, 45 MB L3-cache) and 1000 GB RAM. The machine is running 64-bit Debian 10 with Linux kernel version 4.19.67. Our implementation runs on a single core. For each graph and set of parameters we average the results of ten repetitions.
We use nested dissection in Metis with default parameters. Our reference is Metis without reductions. We also compare our result with orderings from the \texttt{gord}-program from the software package Scotch (version 6.0.6) \cite{scotch}.
In evaluating our orderings we focus on the number of non-zeros in the matrix factors and the running time of the ordering algorithm. 
We obtain the number of non-zeros with the \texttt{gotst}-program from Scotch. This program performs a Cholesky factorization and reports statistics on the elimination process.
Some of our plots are performance profiles. 
These plots relate the running times or quality of all algorithms to the fastest/best algorithm on a per-instance basis.
For each algorithm A, these ratios are sorted in increasing order. The plots show $\big(\frac{t_\text{fastest}}{t_\text{A}}\big)$ (in case of running time) or $\big(\frac{\phi_\text{best}}{\phi_\text{A}}\big)$ on the y-axis. 
A point close to zero shows that the algorithm was considerably slower/worse than the fastest/best algorithm.  \\[-5pt]

\noindent\textbf{Instances.}
We evaluate our algorithm on the large undirected graphs from \cite{meyerhenke14}.
These graphs include social networks, citation networks and web graphs compiled from~\cite{benchmarksfornetworksanalysis} and~\cite{snapnets}. We also use the graphs from Walshaw's graph partitioning archive \cite{soper04}\new{, which are mostly meshes and similar graphs,} and road networks obtained from \cite{dimacschallengegraphpartandcluster}. Properties of our benchmark instances can be found in Table~A.\ref{tab:instances}.\\[-5pt]

\noindent\textbf{Parameters.}
We apply the reductions in a fixed order on each recursion level. The reductions are specified by their first letter; $\Delta$ for triangle contraction. We add a number to the configuration to specify the degree limit on simplicial nodes used for social networks. For example, $SD18$ means simplicial node reduction is applied before degree-2 elimination, with the degree limit set to 18 on the social network dataset.
Note that we never use Reductions \ref{red:path} and \ref{red:degree-2} together. After degree-2 elimination, path compression cannot reduce the graph and degree-2 elimination eliminates any nodes contracted by path-compression. Thus, using all reductions equates to the configuration $SITD\Delta$.

Nodes with high degree can cause simplicial node reduction (Reductions~\ref{red:simplicial}) to be slow. Social networks tend to contain high-degree nodes, so we limit the degree of simplicial nodes on these graphs. On meshes and road networks such nodes do not cause problems. Thus, we do not limit the degree for meshes or road networks. The choice of the degree limit for social networks is discussed in Appendix~\ref{sec:degree-limit}.
We use the default parameters for nested dissection in Metis. For Scotch we choose the default ordering strategy (option \texttt{-cq}), which emphasizes quality over speed.

\subsection{Experimental Results}

\begin{figure}[b!]
\centering
        \input{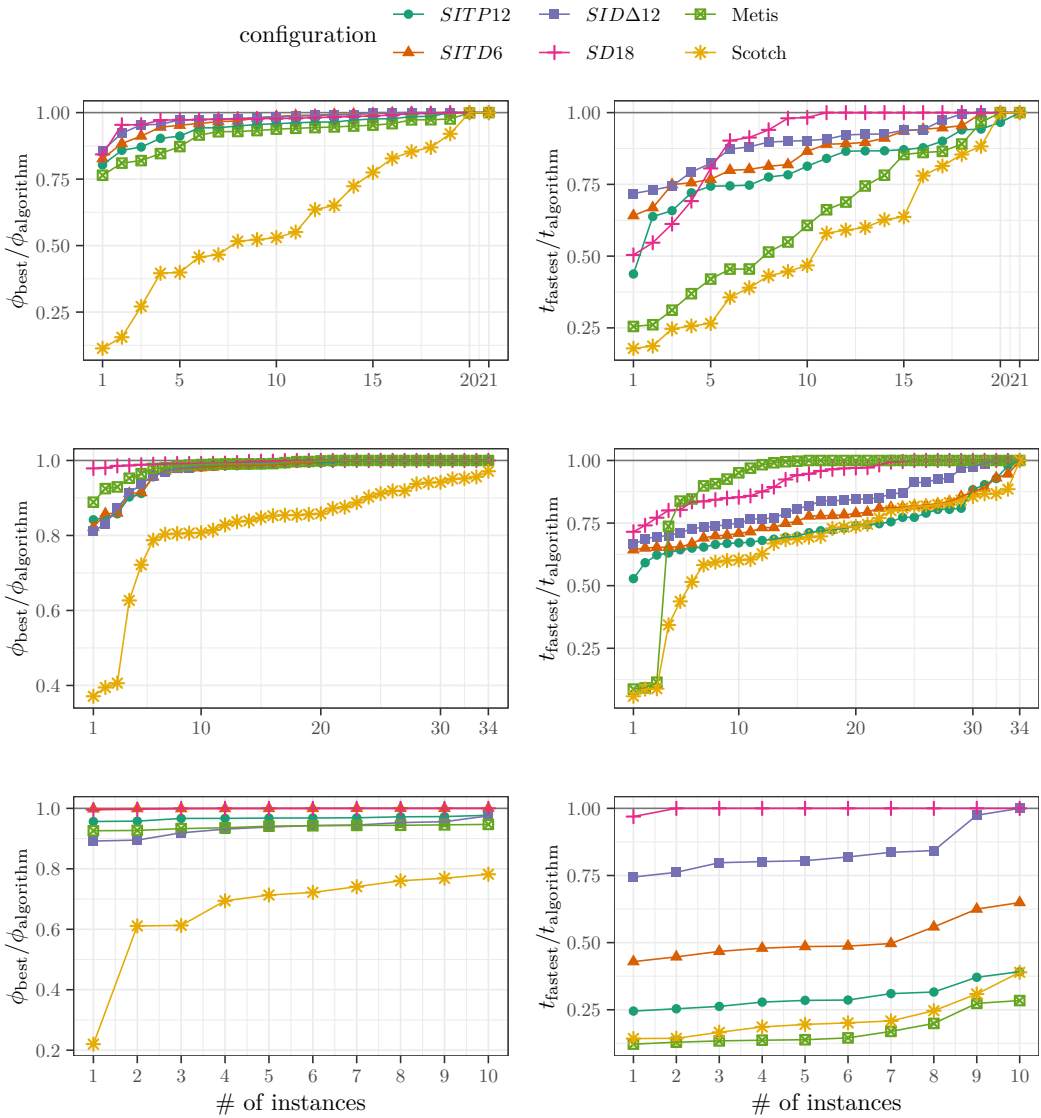}
        \caption{Performance plots for number of non-zeros (left) and running time (right) for different graph classes, from top to bottom: social graphs, meshes and road networks.}
        \label{fig:nnz-performance}
        \label{fig:time-performance}
        \label{fig:road-time-performance}
        \label{fig:road-nnz-performance}
\end{figure}

We now look at the performance of different reductions when used as a preprocessing step before running Metis.
The time reported for our algorithm is the overall running time needed, i.e.,~compute the kernel, run Metis on the kernel, convert the solution on the kernel to a solution on the input graph.
Figure \ref{fig:nnz-performance} compares the results for different combinations of reductions and graph classes for number of non-zeros and running time, respectively. We look at each graph class separately, i.e. social networks, mesh-like networks, and road networks. 
\new{Table~\ref{tbl:per-instance} in Appendix~\ref{sec:per-instance} shows the results for each instance for configuration $SID\Delta12$.}

\textbf{\emph{Social Networks.}} We first look at social networks. In general, reducing the graph before nested dissection yields significant
speedups on most instances over nested dissection without any reductions.
At the same time the number of non-zeros is also reduced.

With configuration $SID\Delta12$ we obtain a speedup of 1.5 on average (see Table \ref{tbl:geomeans}); the improvement in number of non-zeros is 1.06. Note, that for the other configurations, the average speedup is greater than 1.35 on average.
The social networks can be reduced to 57\% of their original size, on average (see Table~\ref{tbl:kernels}).

Out of all graphs and configurations we observe the largest speedup of 3.92 for the instance \texttt{as-22july06}. The smallest speedup for this graph is 1.72 with configuration $SITP12$.
Only two out of 21 of the social graphs do not benefit from the reductions in terms of speedup: on the instances \texttt{eu-2005} and \texttt{as-skitter} nested dissection with reductions is always slower than nested dissection without reductions. For \texttt{as-skitter} the speedup lies between 0.74 and 0.91, for \texttt{eu-2005} between 0.81 and 0.95. 
Out of all graphs and configurations the lowest speedup is 0.75 for instances \texttt{as-skitter} and \texttt{p2p-Gnutella04}, with configuration $SITP12$ in both cases. With configuration $SD18$ we observe a speedup of 1.03 for \texttt{p2p-Gnutella04}. 

The largest improvement in number of non-zeros out of all graphs and configurations is 1.31 relative to Metis for the instance \texttt{coAuthorsCiteseer} with configuration $SITP12$. The speedup is 1.85 for this graph and configuration. 
Only on the instance \texttt{coPapersCiteseer} the number of non-zeros is not reduced when applying reductions. For this graph the number of non-zeros is 4\% above that of Metis with configuration $SD18$. Here, the speedup is 5.43. 
On 14 of the social graphs the number of non-zeros is reduced by all of the configurations. The highest number of non-zeros we observe is $21\%$ higher than that of Metis on the graph \texttt{eu-2005} using configuration $SITP12$.

For this graph class, the largest kernel has 96\% of the nodes of the original graph and is obtained by configuration $SITP12$ for instance \texttt{p2p-Gnutella}.
The smallest kernel has 30\% of the nodes and is obtained by configurations $SITP12$, $SITD12$ and $SID\Delta12$ for instance \texttt{coPapersCiteseer}.

Compared to Scotch and averaged over the social networks our algorithm is between 1.6 and 1.8 times faster than Scotch and produces orderings with an improvement of 1.81 in terms of the number of non-zeros.

\textbf{\emph{Meshes.}} On the meshes, the reductions do not yield a speedup except for a few instances. Those instances are chordal graphs (\texttt{add20}, \texttt{add32}, \texttt{memplus}) and stiffness matrices (\texttt{bcsstk*}).
Chordal graphs are reduced completely by simplicial node reduction. Here, we observe speedups between 5.4 (\texttt{add20}) and 9.9 (\texttt{memplus}).
The stiffness matrices contain many indistinguishable nodes, so the graph size is reduced significantly.
After applying simplicial node reduction and indistinguishable node reduction, \texttt{bcsstk29} is reduced to 72\% of its original size and \texttt{bcsstk30} is reduced to 30\% in terms of number of nodes.
For these stiffness matrices we obtain speedups between 1.04 (\texttt{bcsstk29}) and 2.6 (\texttt{bcsstk32}) with configuration $SID\Delta$. When indistinguishable nodes are not contracted, our algorithm is up to 20\% slower on these instances.
On the other instances the reductions do not have a sufficient impact to reduce running time or number of non-zeros. Only for \texttt{finan512} and \texttt{uk} our algorithm has a speedup greater than 1.01, with 1.09 and 1.04, respectively. No configuration leads to a faster running time on the remaining instances.
The improvement in number of non-zeros ranges from 0.94 (\texttt{vibrobox}, configuration $SID\Delta$) to 1.05 (\texttt{cti}, configurations $SITP$ and $SITD$).
The graphs are reduced by no more than 20\%, on average (see Table~\ref{tbl:kernels}).

Scotch is faster than Metis without reductions on a few instances, but slower in general. Its orderings lead to more non-zeros.
Compared to Scotch, our algorithm is between 1.36 and 1.57 times faster and improves the number of non-zeros between 1.2 and 1.3 times.

\begin{table}[t!]
        \caption{Top: Geometric means of the improvement in number of non-zeros (nnz) relative to Metis (larger is better)  and speedup ($\mathcal{S}$) relative to Metis for different configurations. Bottom: Average number of nodes in the kernel and standard deviation $\sigma$ (smaller is better).}
                 \centering
        \begin{tabular}{lrr@{\hskip 20pt}rr@{\hskip 20pt}rr}
        Graph Class                & \multicolumn{2}{c}{Social} & \multicolumn{2}{c}{Meshes} & \multicolumn{2}{c}{Road} \\
        \midrule
         Reductions                       & \multicolumn{6}{c}{Number of non-zeros} \\
        \midrule
                      & nnz    & $\mathcal{S}$ & nnz                 & $\mathcal{S}$   & nnz  & $\mathcal{S}$              \\
        $SITP12$      & 1.03   & 1.35          & 0.99                & 0.91            & 1.03 & 1.79                       \\
        $SITD6$       & 1.05   & 1.44          & 0.99                & 0.97            & 1.06 & 3.07                       \\
        $SID\Delta12$ & 1.06   & 1.50          & 0.99                & 1.06            & 1.00 & 5.05                       \\
        $SD18$        & 1.06   & 1.49          & 1.01                & 1.06            & 1.06 & 6.03                       \\
        $SD\Delta12$  & 1.05   & 1.44          & 1.01                & 0.99            & 1.00 & 6.37                       \\
        \midrule
                                & \multicolumn{6}{c}{Kernel Sizes} \\
        \midrule
                      & mean   & $\mathcal{\sigma}$ & mean                & $\mathcal{\sigma}$   & mean & $\mathcal{\sigma}$         \\
        $SITP12$      & 0.57   & 0.23               & 0.83                & 0.32                 & 0.37 & 0.18                       \\
        $SITD6$       & 0.58   & 0.22               & 0.82                & 0.32                 & 0.20 & 0.13                       \\
        $SID\Delta12$ & 0.57   & 0.23               & 0.82                & 0.32                 & 0.20 & 0.13                       \\
        $SD18$        & 0.60   & 0.23               & 0.90                & 0.28                 & 0.20 & 0.13                       \\
        $SD\Delta12$  & 0.61   & 0.23               & 0.90                & 0.28                 & 0.20 & 0.13                       \\
        \bottomrule
        \end{tabular}
        \label{tbl:kernels}
        \label{tbl:geomeans}
\end{table}

\textbf{\emph{Road Networks.}} Applying reductions to road networks leads to high speedups (see Figure \ref{fig:road-time-performance}) and improvements in quality (see Figure \ref{fig:road-nnz-performance}). The average speedups are between 1.48 and 6.0. The number of non-zeros is improved between 1.03 and 1.06-fold. %\csch{quantify average first}
Road networks contain many degree-2 nodes, so degree-2 elimination is highly effective. 
After removing simplicial nodes and degree-2 nodes the \texttt{osm} instances retain less than 20\% of their nodes;
the instances \texttt{road\_usa} and \texttt{road\_central} are reduced to around 45\% of their original size.
\new{Simplicial node reduction on its own yields a speedup of 1.35 and an improvement in number of non-zeros by 4\% (see Table~\ref{tbl:road-configs}). Degree-2 elimination without simplicial node reduction does not improve the number of non-zeros, but leads to a 4.69-fold speedup.}
Reducing the road networks by \new{both} simplicial node reduction and degree-2 elimination (configuration $SD$) yields a 6-fold speedup on average (see Table \ref{tbl:geomeans}),
with the lowest speedup at 3.5 and the highest speedup at 8.2. This is also the highest speedup we observe.
The number of non-zeros is improved by 1.06 on average with this configuration.
While triangle contraction further improves the running time, it also leads to a larger number of non-zeros.

Configuration $SITP$ results in the lowest speedups, between 1.3 (\texttt{road\_central}) and 2.2 (\texttt{asia.osm}). With configuration $SID\Delta$ the number of non-zeros is increased on 4 of the 10 road networks, however never by more than 6\%. Configuration $SD$ improves the number of non-zeros the most, by up to 1.07 (\texttt{great-britain.osm}).

On the road networks Scotch is consistently faster than Metis without reductions, but the quality of its orderings is significantly worse.
Compared to Scotch, our algorithm is between 1.4 and 5 times faster whenever degree-2 elimination or path compression are used, on average. Otherwise, our algorithm is slower. The number of non-zeros is always improved, between 1.5 and 1.6 times.

\textbf{\emph{Using All Reductions.}}
The configuration $SITD\Delta$ uses all reductions. For this configuration degree limit 12 results in the best performance.
For all graph classes, using all reductions is no better than using configuration $SID\Delta12$. The kernels obtained by the former are within 1\% of the size of the kernels obtained by the latter, on average. This is not sufficient to reduce  the running time. On the social networks, the speedup of configuration $SITD\Delta12$ is 1.44, on average, which is lower than the speedup of 1.5 obtained with configuration $SID\Delta12$. On the meshes, the speedup of configuration $SITD\Delta$ is 0.94; on the road networks it is 4.11, on average.
The improvement in number of non-zeros does not change by more than 1.5\% between the two configurations.

Adding triangle contraction to the configuration $SITD$ yields the configuration $SITD\Delta$.
With the configuration $SITD\Delta12$ we achieve a speedup of 4.11 on the road networks. The number of non-zeros is reduced compared to Metis, the improvement being 0.99. On the social networks, the average speedup does not change and the improvement in number of non-zeros is reduced by less than 1\%. On the meshes the number of non-zeros is not changed and the running time is reduced, with the average speedup at 0.94. Adding triangle contraction to configuration $SITD$ does not lead to an improvement in running time or quality. On road networks we get a faster running time at the expense of quality.

\begin{table}[t!]
        \centering
        \caption{Average speedup $\mathcal{S}$, improvement in number of non-zeros (nnz) and kernel size $n'$ from simplicial node reduction and degree-2 elimination on the road networks.}
        \label{tbl:road-configs}
        \begin{tabular}{lrrr}
                \toprule
                Configuration   & $\mathcal{S}$ & nnz  & $n'$ \\
                \midrule
                $S$             & 1.04          & 1.35 & 0.77 \\
                $D$             & 1.00          & 4.69 & 0.30 \\
                $SD$            & 1.06          & 6.03 & 0.20 \\
                \bottomrule
        \end{tabular}
                \vspace*{-.25cm}
\end{table}

\section{Conclusion}
\label{s:conclusion}

\new{By applying data reduction rules exhaustively we obtain improved quality \emph{and} at the same time large improvements in running time on a variety of instances.
Overall, we arrive at a system that outperforms the state-of-the-art significantly.

On road networks we obtain orderings with lower fill-in six times faster than nested dissection alone. As orderings of such networks are used in preprocessing of shortest path algorithm like customizable contraction hierarchies, we believe that the additional reductions presented here can yield a significant speed up in the preprocessing time of such algorithms~\cite{DBLP:journals/jea/DibbeltSW16,DBLP:journals/algorithms/GottesburenHUW19}.}

We have so far not explored the use of these reduction rules in combinations with other algorithms for the minimum fill-in problem. However, the rules presented here are mostly independent of the underlying algorithm. In particular, eliminating simplicial nodes whenever possible appears to be very effective in reducing running time without harming the quality of the resulting ordering. \new{Other important future work includes parallelization. Given the good results, we plan to release our software.}

\bibliographystyle{myplainnat}
\bibliography{phdthesiscs}

\begin{appendix}
\section{Instances}
\begin{table*}[h!]
        \centering
        \caption{Basic properties of the graphs.}
        \label{tab:instances}
        \begin{tabular}{lrr|lrr}
\toprule
graph                   & $|V|$  & $|E|$    & graph                       & $|V|$   & $|E|$    \\
\midrule
\multicolumn{6}{l}{Social Networks} \\
\midrule
amazon-2008       & \numprint{735323} & \numprint{3523472}  & eu-2005               & \numprint{862664}  & \numprint{16138468} \\
as-22july06       & \numprint{22963}  & \numprint{48436}    & in-2004               & \numprint{1382908} & \numprint{13591473} \\
as-skitter        & \numprint{554930} & \numprint{5797663}  & loc-brightkite\_edges & \numprint{56739}   & \numprint{212945}   \\
citationCiteseer  & \numprint{268495} & \numprint{1156647}  & loc-gowalla\_edges    & \numprint{196591}  & \numprint{950327}   \\
cnr-2000          & \numprint{325557} & \numprint{2738969}  & p2p-Gnutella04        & \numprint{6405}    & \numprint{29215}    \\
coAuthorsCiteseer & \numprint{227320} & \numprint{814134}   & PGPgiantcompo         & \numprint{10680}   & \numprint{24316}    \\
coAuthorsDBLP     & \numprint{299067} & \numprint{977676}   & soc-Slashdot0902      & \numprint{28550}   & \numprint{379445}   \\
coPapersCiteseer  & \numprint{434102} & \numprint{16036720} & web-Google            & \numprint{356648}  & \numprint{2093324}  \\
coPapersDBLP      & \numprint{540486} & \numprint{15245729} & wiki-Talk             & \numprint{232314}  & \numprint{1458806}  \\
email-EuAll       & \numprint{16805}  & \numprint{60260}    & wordassociation-2011  & \numprint{10617}   & \numprint{63788}    \\
enron             & \numprint{69244}  & \numprint{254449}   &                       &      &          \\ 
\midrule
\multicolumn{6}{l}{Walshaw Benchmark} \\
\midrule
144      & \numprint{144649} & \numprint{1074393} & fe\_4elt2   & \numprint{11143}  & \numprint{32818}   \\
3elt     & \numprint{4720}   & \numprint{13722}   & fe\_body    & \numprint{45087}  & \numprint{163734}  \\
4elt     & \numprint{15606}  & \numprint{45878}   & fe\_ocean   & \numprint{143437} & \numprint{409593}  \\
598a     & \numprint{110971} & \numprint{741934}  & fe\_pwt     & \numprint{36519}  & \numprint{144794}  \\
add20    & \numprint{2395}   & \numprint{7462}    & fe\_rotor   & \numprint{99617}  & \numprint{662431}  \\
add32    & \numprint{4960}   & \numprint{9462}    & fe\_sphere  & \numprint{16386}  & \numprint{49152}   \\
auto     & \numprint{448695} & \numprint{3314611} & fe\_tooth   & \numprint{78136}  & \numprint{452591}  \\
bcsstk29 & \numprint{13992}  & \numprint{302748}  & finan512    & \numprint{74752}  & \numprint{261120}  \\
bcsstk30 & \numprint{28924}  & \numprint{1007284} & m14b        & \numprint{214765} & \numprint{1679018} \\
bcsstk31 & \numprint{35588}  & \numprint{572914}  & memplus     & \numprint{17758}  & \numprint{54196}   \\
bcsstk32 & \numprint{44609}  & \numprint{985046}  & t60k        & \numprint{60005}  & \numprint{89440}   \\
bcsstk33 & \numprint{8738}   & \numprint{291583}  & uk          & \numprint{4824}   & \numprint{6837}    \\
brack2   & \numprint{62631}  & \numprint{366559}  & vibrobox    & \numprint{12328}  & \numprint{165250}  \\
crack    & \numprint{10240}  & \numprint{30380}   & wave        & \numprint{156317} & \numprint{1059331} \\
cs4      & \numprint{22499}  & \numprint{43858}   & whitaker3   & \numprint{9800}   & \numprint{28989}   \\
cti      & \numprint{16840}  & \numprint{48232}   & wing        & \numprint{62032}  & \numprint{121544}  \\
data     & \numprint{2851}   & \numprint{15093}   & wing\_nodal & \numprint{10937}  & \numprint{75488}   \\
\midrule
\multicolumn{6}{l}{Road Networks} \\
\midrule

asia.osm          & \numprint{11950757} & \numprint{12711603} & italy.osm         & \numprint{6686493}  & \numprint{7013978}  \\
belgium.osm       & \numprint{1441295}  & \numprint{1549970}  & luxembourg.osm    & \numprint{114599}   & \numprint{119666}   \\
europe.osm        & \numprint{50912018} & \numprint{54054660} & netherlands.osm   & \numprint{2216688}  & \numprint{2441238}  \\
germany.osm       & \numprint{11548845} & \numprint{12369181} & road\_central     & \numprint{14081816} & \numprint{16933413} \\
great-britain.osm & \numprint{7733822}  & \numprint{8156517}  & road\_usa         & \numprint{23947347} & \numprint{28854312} \\

\bottomrule
        \end{tabular}
\end{table*}
\vfill
\pagebreak

\section{Per-Instance Results}
\label{sec:per-instance}

\begin{table}[H]
\centering
\caption{Speedup $\mathcal{S}$ relative to Metis (larger is better), improvement in number of non-zeros NNZ relative to Metis (larger is better) and number of nodes in the kernel $n'$ (smaller is better) per instance for configuration $SID\Delta12$. \new{Note that \texttt{add32} is a chordal graph, but is not reduced completely by simplicial node reduction due to the details of our implementation.}}
\label{tbl:per-instance}
\begin{tabular}{lrrr|lrrr}
\toprule
graph                 & $\mathcal{S}$ & nnz & $n'$ & $\mathcal{S}$ & nnz & $n'$ \\
\midrule
\multicolumn{3}{l}{Social Networks} \\
\midrule

amazon-2008           & 1.01 & 1.02 & 0.80 & eu-2005               & 0.93 & 1.01 & 0.84 \\
as-22july06           & 2.82 & 1.06 & 0.44 & in-2004               & 1.50 & 1.00 & 0.62 \\
as-skitter            & 0.91 & 1.05 & 0.91 & loc-brightkite\_edges & 1.55 & 1.07 & 0.51 \\
citationCiteseer      & 1.13 & 1.04 & 0.79 & loc-gowalla\_edges    & 1.34 & 1.03 & 0.61 \\
cnr-2000              & 1.20 & 1.01 & 0.60 & p2p-Gnutella04        & 0.93 & 1.01 & 0.93 \\
coAuthorsCiteseer     & 2.20 & 1.27 & 0.30 & PGPgiantcompo         & 2.83 & 1.04 & 0.29 \\
coAuthorsDBLP         & 2.20 & 1.22 & 0.31 & soc-Slashdot0902      & 0.86 & 1.02 & 0.90 \\
coPapersCiteseer      & 1.77 & 0.86 & 0.29 & web-Google            & 1.28 & 1.09 & 0.72 \\
coPapersDBLP          & 1.79 & 1.08 & 0.43 & wiki-Talk             & 1.01 & 1.06 & 0.50 \\
email-EuAll           & 3.60 & 1.15 & 0.25 & wordassociation-2011  & 1.30 & 1.21 & 0.60 \\
enron                 & 2.50 & 1.05 & 0.29 &                       &      &      &      \\

\midrule
\multicolumn{3}{l}{Meshes} \\
\midrule

144                   & 0.77 & 0.99 & 1.00 & fe\_4elt2             & 0.75 & 1.00 & 1.00 \\
3elt                  & 0.67 & 1.00 & 1.00 & fe\_body              & 0.82 & 1.00 & 0.96 \\
4elt                  & 0.69 & 0.99 & 1.00 & fe\_ocean             & 0.89 & 0.99 & 1.00 \\
598a                  & 0.78 & 0.99 & 1.00 & fe\_pwt               & 0.77 & 1.00 & 1.00 \\
add20                 & 5.69 & 1.15 & 0.00 & fe\_rotor             & 0.72 & 1.00 & 1.00 \\
add32                 & 7.19 & 1.05 & 0.01 & fe\_sphere            & 0.83 & 1.00 & 1.00 \\
auto                  & 0.69 & 0.99 & 1.00 & fe\_tooth             & 0.71 & 1.01 & 1.00 \\
bcsstk29              & 1.05 & 0.92 & 0.73 & finan512              & 0.94 & 0.97 & 0.86 \\
bcsstk30              & 2.34 & 0.85 & 0.33 & m14b                  & 0.74 & 1.01 & 1.00 \\
bcsstk31              & 1.77 & 0.85 & 0.49 & memplus               & 9.40 & 1.09 & 0.00 \\
bcsstk32              & 2.64 & 0.88 & 0.33 & t60k                  & 0.84 & 1.01 & 0.98 \\
bcsstk33              & 1.53 & 0.88 & 0.50 & uk                    & 0.80 & 0.97 & 0.86 \\
brack2                & 0.78 & 0.97 & 1.00 & vibrobox              & 0.72 & 0.94 & 0.99 \\
crack                 & 0.79 & 0.99 & 1.00 & wave                  & 0.83 & 1.04 & 1.00 \\
cs4                   & 0.80 & 1.02 & 1.00 & whitaker3             & 0.62 & 0.99 & 1.00 \\
cti                   & 0.84 & 1.04 & 1.00 & wing                  & 0.80 & 0.99 & 1.00 \\
data                  & 0.73 & 1.01 & 1.00 & wing\_nodal           & 0.77 & 1.02 & 1.00 \\

\midrule
\multicolumn{3}{l}{Road Networks} \\
\midrule

asia.osm              & 5.8 & 1.05 & 0.14 & italy.osm              & 6.2 & 1.01 & 0.11 \\
belgium.osm           & 4.5 & 0.98 & 0.16 & luxembourg.osm         & 5.4 & 1.02 & 0.10 \\
europe.osm            & 6.2 & 1.01 & 0.14 & netherlands.osm        & 4.2 & 1.01 & 0.22 \\
germany.osm           & 5.5 & 0.99 & 0.16 & road\_central          & 3.4 & 0.95 & 0.44 \\
great-britain.osm     & 6.7 & 1.01 & 0.14 & road\_usa              & 3.7 & 0.94 & 0.45 \\

\bottomrule

\end{tabular}
\end{table}
\vfill
\pagebreak

\section{Choice of Degree Limit for Simplicial Node Reduction}
\label{sec:degree-limit}

\begin{figure}[H]
        \centering
        \input{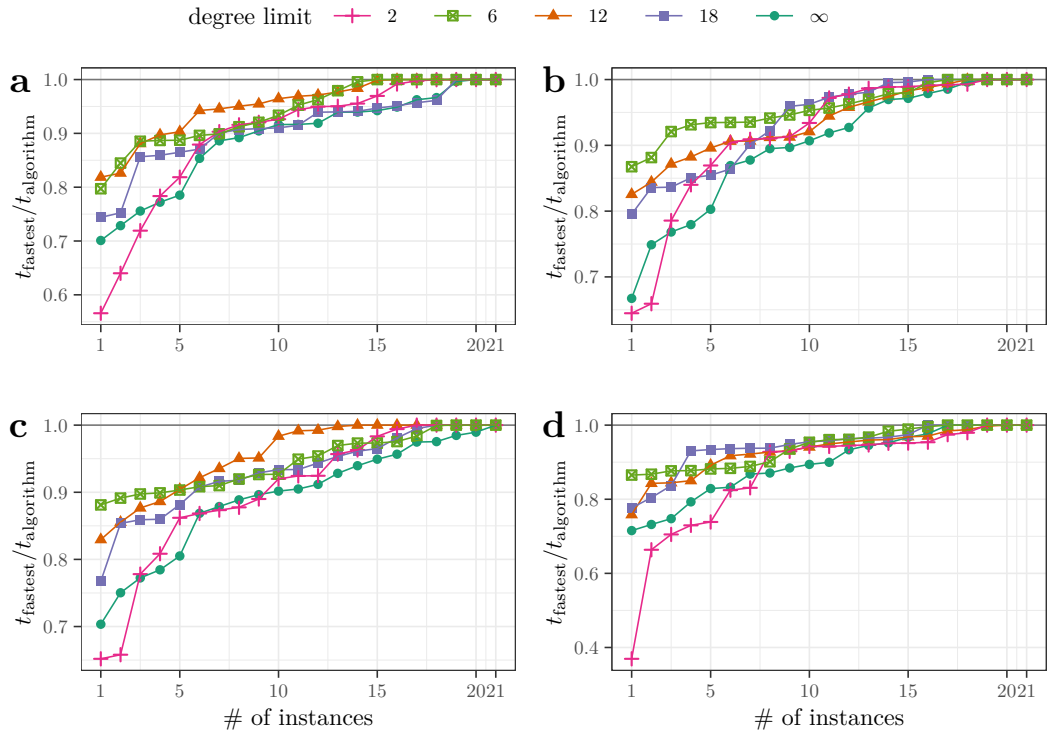}
        \caption{Running time performance of different degree limits for four configurations: $SITP$ (\textbf{a}), $SITD$ (\textbf{b}), $SID\Delta$ (\textbf{c}), $SD$ (\textbf{d}).}
        \label{fig:snw-degree-limit}
\end{figure}

Figure~\ref{fig:snw-degree-limit} shows how the degree limit for simplicial nodes influences the running time for the configurations $SITP$, $SITD$, $SID\Delta$ and $SD$. Based on this we choose a limit of 12 for configurations $SITP$ and $SID\Delta$, 6 for $SITD$ and 18 for $SD$.

{}

\end{appendix}

\end{document}